\documentclass[12pt, a4paper]{iopart}

\pdfoutput=1
\usepackage{graphicx}
\usepackage{hyperref}
\hypersetup{colorlinks=true,%
            urlcolor=blue,%
            linkcolor=red,%
            citecolor=red}

\begin{document}
\title{Effective Administration of a Large Undergraduate Physics Class: From
Enrolment to Assessment Feedback}%
\author{MM~Casey and MW~Kille}
\date{Draft: \today}
\address{SUPA, School of Physics \& Astronomy, University of Glasgow, GLASGOW
{G12~8QQ}, UK.}
\ead{\href{mailto:Morag.Casey@glasgow.ac.uk}{Morag.Casey@glasgow.ac.uk}}
\begin{abstract}
At the beginning of academic year 2007-08, staff in the School of Physics \&
Astronomy at the University of Glasgow started to implement a number of
substantial changes to the first year physics class. The main aims were to
improve the academic performance and progression statistics for the class, with
early identification of ``at-risk'' students being one priority. The introduction of
novel electronic and computer-based data processing strategies ensured both a
quick turnaround time in processing student records and also a much-improved
confidence in the accuracy of assessment feedback returned to students. These
techniques were a contributory factor in helping raise the pass-rate by $\sim10\%$
and the direct progression rate by $\sim13\%$ by the end of 2008-09. The
effectiveness of these processes have proved adequate for coping with the
unexpected and dramatic 50\% increase in class size 2009-10 and are now in the
process of being rolled out to other classes within the School.

\vspace{0.25cm}
\end{abstract}

\vspace{2cm}
\noindent{\it Journal of Science Education / Revista de Educaci\'{o}n en
Ciencias
\bf{12 (2)}, 2011}

\newpage
\title[Effective Administration of a Large Undergraduate Physics
Class]{Administraci\'{o}n Efectiva de un Grupo Grande de la Licenciatura en
F\'{i}sica: Desde la Inscripci\'{o}n hasta la
Retroalimentaci\'{o}n de Evaluaciones}%
\author{MM~Casey and MW~Kille}
\date{Draft: \today}
\address{SUPA, School of Physics \& Astronomy, University of Glasgow, GLASGOW
{G12~8QQ}, UK.}
\ead{\href{mailto:Morag.Casey@glasgow.ac.uk}{Morag.Casey@glasgow.ac.uk}}
\begin{abstract}
Al comienzo del a\~{n}o acad\'{e}mico 2007-08, el personal en el Departamento de
F\'{i}sica y Astronom\'{i}a en la Universidad de Glasgow empez\'{o} a
implementar una serie de cambios sustanciales al grupo de f\'{i}sica de primer
a\~{n}o. Los principales objetivos fueron el mejorar el rendimiento
acad\'{e}mico y el ascenso estad\'{i}stico del grupo, con la identificaci\'{o}n
temprana de estudiantes en riesgo como una primera prioridad. La
introducci\'{o}n de nuevas estrategias electr\'{o}nicas y de procesamiento de
datos basados en computadoras, aseguraron un tiempo r\'{a}pido de entrega de
notas a los estudiantes, as\'{i}  como una muy mejorada confianza en la
precisi\'{o}n de la retroalimentaci\'{o}n de las evaluaciones hacia ellos.
\'{E}stas t\'{e}cnicas fueron un factor que contribuy\'{o} en el incremento de
la raz\'{o}n de pase en aproximadamente un 10\% y en la raz\'{o}n de progression
directa en aproximadamente 13\% hacia el final de 2008-09.  La efectividad de
estos procesos ha probado adecuadamente como enfrentarse al inesperado y
dram\'{a}tico incremento del 50\% en el tama\~{n}o del grupo en 2009-10 por lo
que \'{e}stos est\'{a}n ahora siendo aplicados en otros grupos dentro del
Departamento.

\vspace{0.25cm}
\noindent{\it Palabras Claves\/}: licenciatura, estructuraci\'{o}n, administraci\'{o}n, 
retenci\'{o}n, asistencia
\end{abstract}

\vspace{2cm}
\noindent{\it Journal of Science Education / Revista de Educaci\'{o}n en
Ciencias \bf{12 (2)}, 2011}

\maketitle

\section{Introduction}
Retention and completion rates for first year classes have been an area of
concern in UK universities for some time~\cite{crosling}. Large first year
class sizes have traditionally presented challenges for staff wishing to
identify and track students who have disengaged from the process of studying;
first year physics at the University of Glasgow is no different. In 2007-08,
staff initiated a number of substantial changes to the ways in which academic
assessment and personal support were offered to students~\cite{casey2007,casey2009}. 
In order to manage this effectively, concerted effort was also
directed towards the development of computer-based data processing strategies. 

The use of computer-based technologies to administer large classes is not a new
idea~\cite{simair, visscher} nor is the idea of
using such technologies to track and support students classified as being
``at-risk''~\cite{middleton}. In this case, however, increasing student numbers
as well as an increase in the fraction of continuous assessment required for
credit in the class demanded that all existing administrative processes be
streamlined. The purpose of this paper is to present an overview of the
development of those strategies as well as a review of their effectiveness.

\section{First year Physics at the University of Glasgow}
The context of first year physics within the wider structure of undergraduate
degrees both at the University of Glasgow and within the Scottish Educational
System is described elsewhere~\cite{casey2007,casey2009}. In summary, first year
science students undertake a general curriculum with physics making up one third
of that year's credits; typically 50\% of those students do not initially intend
to continue with physics beyond first year.

The main part of the academic year at the University of Glasgow runs from
September to March with two 11-week semesters~\cite{ug}. The majority of course material
in first year physics is delivered via full-class lectures although laboratory
classes play an important role. Assessment is weighted towards the end-of-year
examination (60\% of the final grade) but continuous assessment plays an
important role throughout the year in the form of regular workshop-tests (20\% of
final grade) and laboratory classes (20\% of the final grade). The main peaks of
activity in terms of data-processing are: enrolment (mid-September to
mid-October); lecture attendance monitoring (daily); laboratory attendance
monitoring (daily) and workshop-test attendance monitoring and marking (four per
semester)

When the workshop-tests were introduced in 2007-08 in a class of $\sim160$\,students,
it became impractical to continue to process records manually and the decision
to introduce computer-based technologies was made. A minimum of 50\% attendance
at laboratory classes and workshop-tests is required for credit in first year
physics. Low levels of engagement, academic performance and retention have been
found to be correlated with low attendance at lectures~\cite{maloney, thatcher}
and a decision was made to monitor lecture attendance too.

\section{New Technologies and Processes for 2007-08}
In order to process data by computer, {\it form recognition} software~\cite{autonomy}
and portable barcode scanner technologies~\cite{opticon} were
introduced to the administration of the class. {\it Form recognition} software is
capable of interpreting hand-written letters and numbers by means of {\it optical
character recognition (OCR)} as well as reading shaded ``bubbles'' and ``ticked
boxes'' on forms. Thus, it is particularly useful for collating the answers from
{\it multiple choice question ({\it MCQ})} papers.

\subsection{Enrolment}
\begin{figure}[ht!]
\begin{center}
\includegraphics[width=.75\textwidth]{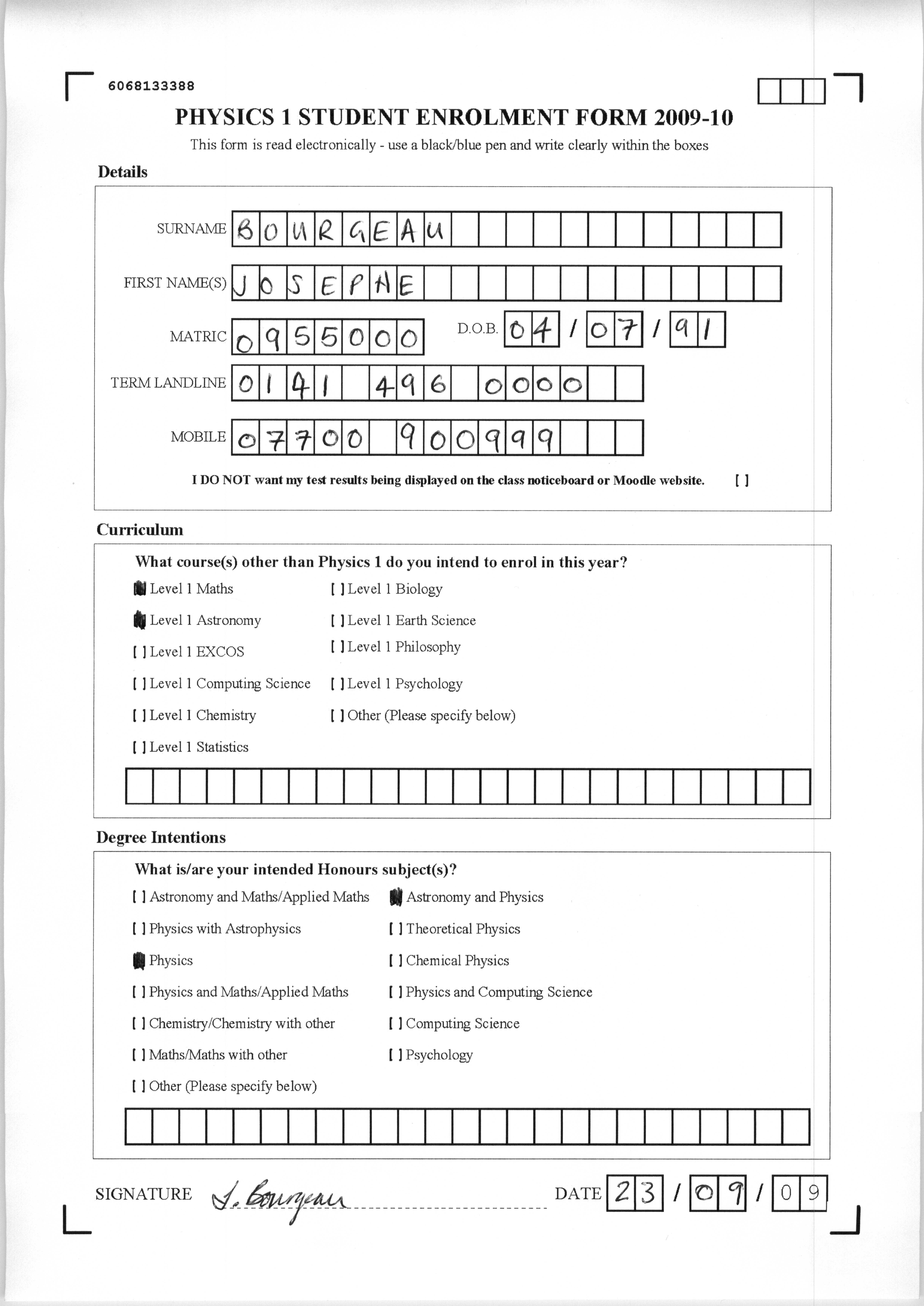}
\caption{Paper enrolment form used for the first year physics class.}
\label{fig1}
\end{center}
\end{figure}

Students wishing to take first year physics must enrol in the class in two
different ways: formally in the University records system ({\it Websurf}) and,
separately, by filling out a paper enrolment form at the first class meeting.
The paper enrolment form (Figure~\ref{fig1}) is designed for {\it form recognition},
batch-processed by a high-speed document scanner~\cite{canon} and the data
therein downloaded into a bespoke {\it MySQL}~\cite{sun} database from
which all post-processing originates.

The information requested in the enrolment form may appear minimal and, indeed,
consists primarily of data used for internal statistical purposes such as
predicting and tracking progression rates (``Curriculum'' and ``Degree Intentions''
sections). Student addresses are not requested because this information already
exists in the University's student records system, {\it Websurf}, but mobile telephone
numbers are important as they can be used to contact students by Short Message
Service (SMS)~\cite{casey2007}. To conform to the UK Government's Data Protection
Act~\cite{ukgov}, students are requested to object to having their
continuous assessment results published on the School's lecture theatre notice
boards.

Although it may appear a straightforward task to ask students to enrol in two
separate ways, it is one which is presents a number of pitfalls. At the start of
the academic year, students often change classes and degree courses after only a
few days attendance; frequently they do this without informing staff.
Consequently, the class list contained in the central records system, {\it Websurf},
may not reflect the local class list held by the School. It is not always
possible to differentiate between students who have enrolled in {\it Websurf} but have
dropped the class without informing a member of staff and students who have
enrolled in {\it Websurf} but are simply not attending class meetings. 

\begin{figure}[ht!]
\begin{center}
\includegraphics[width=.4\textwidth]{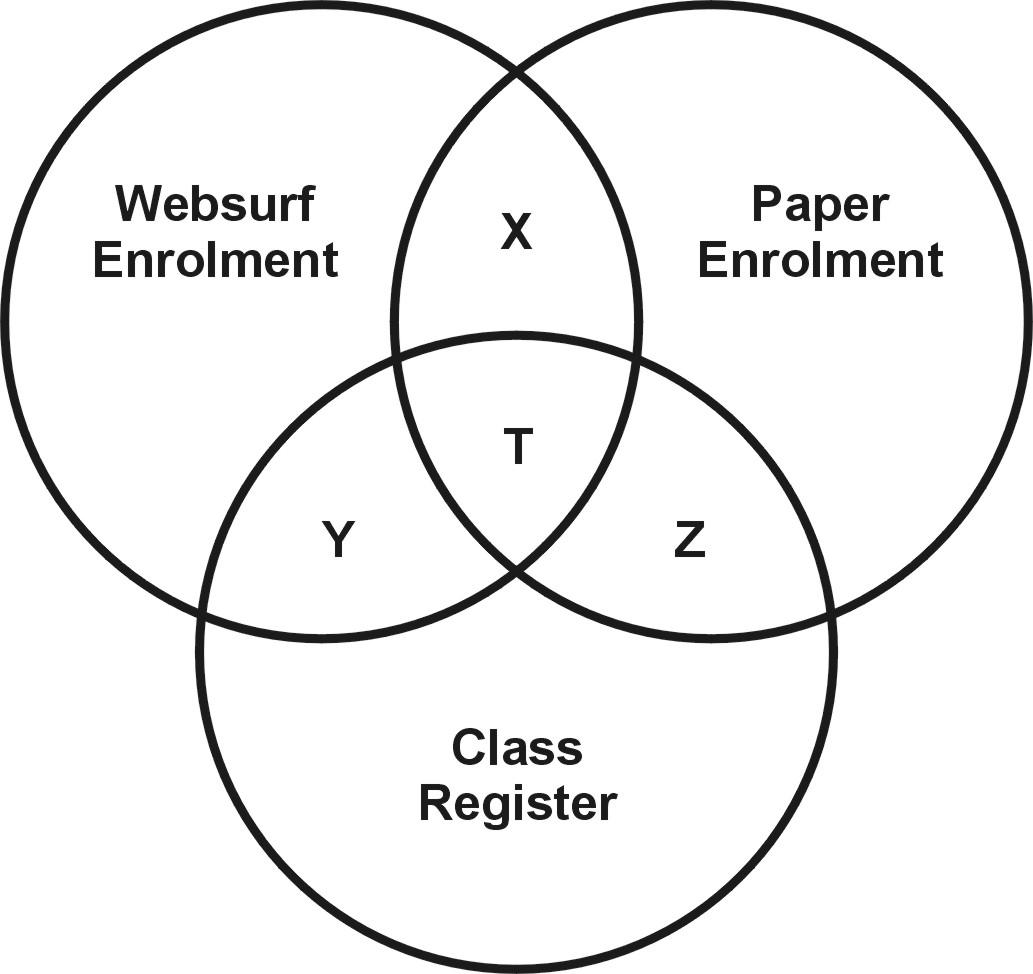}
\caption{Venn diagram showing the relationships between the University records
system, ``{\it Websurf}'', the paper enrolment forms used in the first year physics
class and the class register.}
\label{fig2}
\end{center}
\end{figure}

Effective monitoring of attendance and associated work submission stems from
having an accurate class register and the problem in finalizing the class
register can be illustrated graphically (Figure~\ref{fig2}). At approximately 3 weeks
into the semester (when it is assumed that the flux in student numbers has
fallen effectively to zero), a snapshot from the {\it Websurf} list is compared to the
paper enrolment forms and the class register is initially compiled from both
lists. However, because those two sets can be mutually exclusive, regions Y, Z
and T signify varying degrees of confidence in the attendance data collected for
any class component. 

The problem in administering the class is that it is not efficient to spend time
chasing apparently absent students who have departed the class without telling
anyone. Such students would not be expected to attend but it is not always
possible to tell them apart from students who have not dropped the class but who
have been absent. When analysing the attendance data, there is maximum
confidence for students whose names appear in region T. There is a lack of
confidence in determining the absence of students whose names appear in regions
Y or Z because they may have dropped the class without informing staff. Region
X, however, signifies a genuine absence from someone who would be expected to be
present.

The intention is to minimize the amount of time spent chasing non-existent
students. Therefore it is desirable to force an overlap between the data from
the {\it Websurf} and {\it Paper Enrolment} sets since that maximizes confidence in
correctly contacting ``at-risk'' students. Experience has shown that this requires
the application of personal contact with students.

\subsection{Lecture Attendance Monitoring}
All students at the University of Glasgow are issued registration cards
(formally known as matriculation cards) at the start of each academic session.
This identification card has a unique barcode related to the student's
registration number printed on the front which can be read by any commercial
barcode scanner. It would be possible to monitor attendance at lectures or
laboratory classes by installing fixed barcodes scanner at the entrance to all
rooms. Prior to the current system being introduced, a barcode scanner system
was installed at the entrance to one lecture room. However, this resulted in
severe bottle-necking as large numbers of students tried to enter or leave the
room in the 10-minute switchover period between lectures. Therefore, the
decision was taken to purchase several portable hand-held barcode 
scanners~\cite{opticon}. Students were asked to pass these round the
lecture theatre during the lecture itself, a quiet beep from the scanner
indicating success in scanning the registration card's barcode. 

The use of portable hand-held barcode scanners also makes the system adaptable
to future changes; if a lecture is moved at short notice to another room or,
even, another building, a lecturer does not have to rely on the installation of
fixed scanners in that new room; the portable scanner can be taken with them.
More pertinently, the system has already been adapted for use in other classes
within the School (such as second year physics) by purchasing, at low additional
cost, extra handsets with the use of multiple handsets eliminating ambiguity in
the data from different classes.

{\it A priori}, staff opinion was that the beeping would be disruptive to lectures (it
was less disruptive than students talking quietly), that the scanners would be
removed by the students (this has never happened) or that the students would
resent being asked to record their attendance every day (in fact, students
appear entertained by idea of scanning their registration cards). The worst that
has happened so far is that students, as well as scanning their registration
cards, sometimes also scan the bar codes on junk-food packaging. However the
barcode numbers on junk-food packaging are sufficiently different to those on
the University of Glasgow's registration cards so can be easily filtered out of
the data-stream. 

The data from the barcode scanner is downloaded into a {\it MySQL} database on a daily
basis and post-processing allows the profiling of whole-class or individual
attendance trends as required. The provision of an accurate class register
ensures that, first and foremost, any non-attending students can be identified
and, subsequently, contacted to ascertain if they would benefit from any extra
support. Post-processing of the whole-year lecture attendance data may also be
undertaken to analyse longitudinal trends.

\subsection{Laboratory Attendance Monitoring}
Initially, it was believed that the lecture attendance model described above
could be applied wholesale to the laboratory classes. However, the laboratory
classes are often subdivided into several rooms which made identification of
individual students problematic. Therefore, a hybrid system where students
signed a paper register was introduced.

\begin{figure}[ht!]
\begin{center}
\includegraphics[width=.75\textwidth]{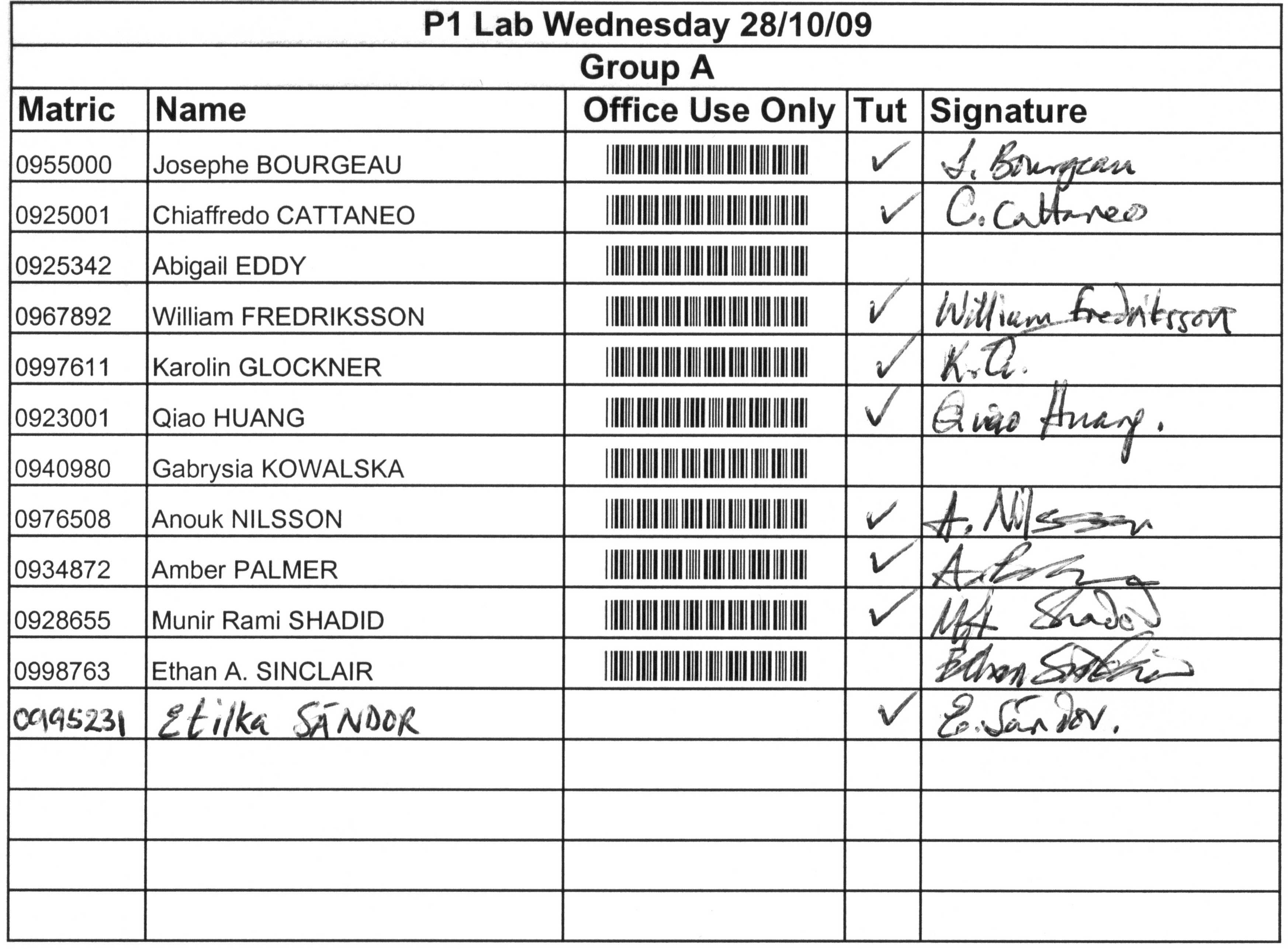}
\caption{Example of a first-year laboratory attendance register. }
\label{fig3}
\end{center}
\end{figure}

A typical paper laboratory register (Figure~\ref{fig3}) consists of the list of students
assigned to a particular experiment on a particular day of the week. As well as
the usual name, registration number and signature there is a barcode version of
the student's registration number. Students sign against their name on the list
and a manual scan of those barcodes with signatures next to them will provide a
list of those who are present in the class.  The column marked ‘Tut’ is used for
noting which students have handed in their pre-laboratory tutorial questions.

The staff member in charge of the laboratory class that day has responsibility
for asking students to sign the register and ensuring that the number of
signatures matches the number of bodies in the room. Because students are
allowed some flexibility in attending the laboratory on a day different to their
nominated day or, for that matter, twice in one week to make up a missed class,
they are allowed to sign their name at the end of the main list. A cross-check
of the number of signatures against the number of barcodes downloaded to the
database subsequently reduces the margin for false positives in attendance to
zero. 

By the start of 2009-10, there were 240 students in the class but the job of
processing the from students housed in up to three separate rooms on five
separate laboratory days was refined enough that the task would typically be
finished within 30-minutes to one hour of the start of any laboratory class.
This meant that absent students could be contacted very quickly. In fact, during
the first week of laboratory classes, the policy was to contact students by SMS
within one hour of the laboratory class starting every day. After the first week
of laboratory classes, the data-processing frequency was reduced to once-weekly
to reduce staff effort. 

\subsection{Workshop-test Attendance and Marking}
\begin{figure}[ht!]
\begin{center}
\includegraphics[width=.75\textwidth]{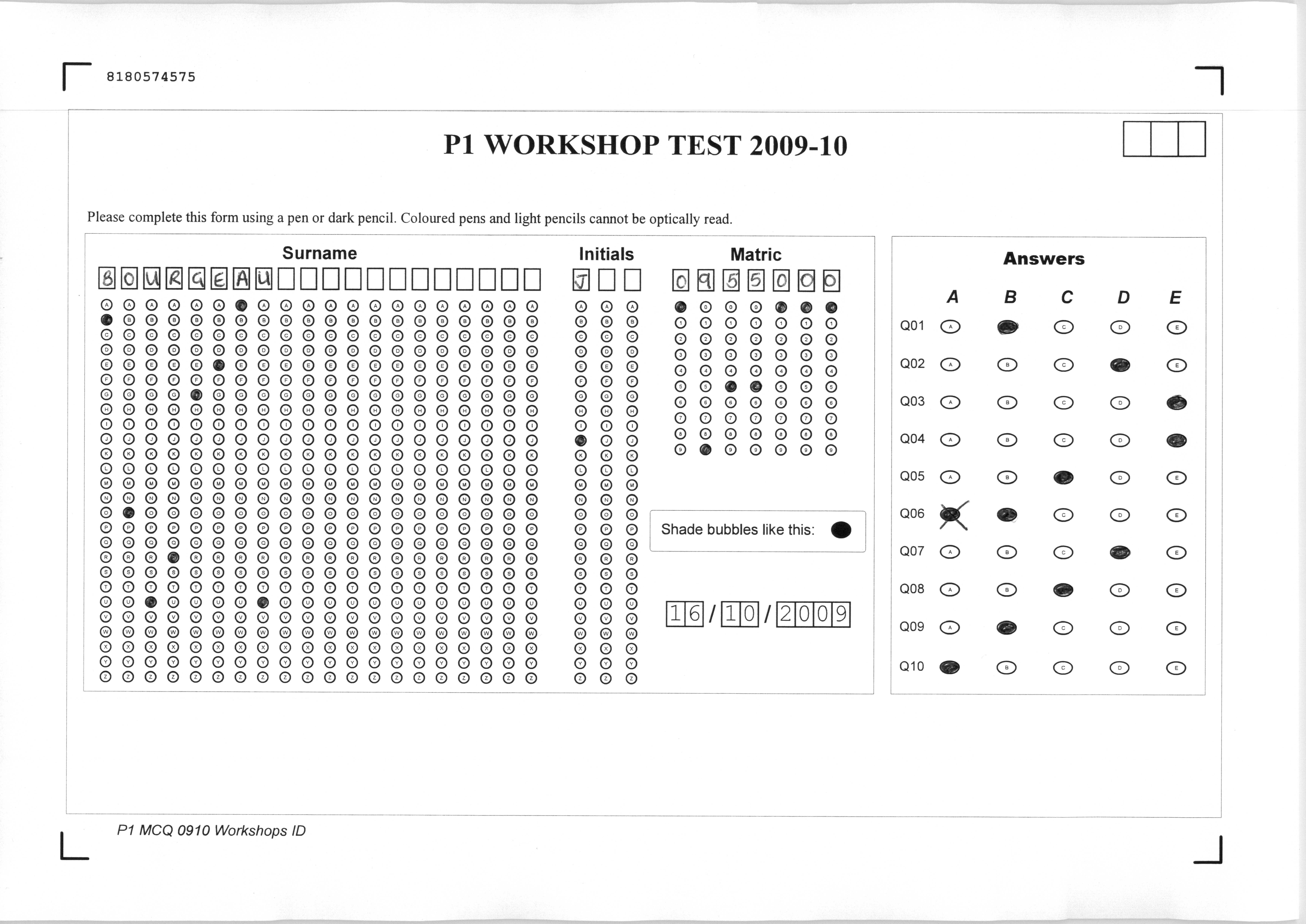}
\caption{Example of a first year physics workshop-test answer sheet.}
\label{fig4}
\end{center}
\end{figure}

In 2007-08, the class head introduced a series of regular multiple choice
question tests ({\it MCQ}) as a means of increasing the fraction of continuous
assessment attributed to the first year physics course. It was desirable to mark
these tests quickly, accurately and with a minimum of staff effort. Therefore,
computer-based marking was implemented with {\it form recognition} technology used to
read the answer-sheets (Figure~\ref{fig4}) and log the answers into the database for
attendance monitoring as well as test-marking.

 Students shade in the circles corresponding to the letters in their surname and
numbers corresponding to their matriculation number. Although they can also
write their name and matriculation number in the boxes above, it is the shaded
circles that are read by the {\it form recognition} software. Students fill in their
answers to the multiple-choice questions via the panel on the right-hand side.
If they change their mind about the answer they wish to submit (as shown in
question Q06), they may put a cross through the old answer and shade in a new
answer. In such cases, the computer software used to scan the answer sheet
generates a request for manual intervention to resolve the conflict. 

\subsubsection{Workshop-test Attendance}
Attendance at the workshop-tests was checked by a series of measures. The answer
sheets themselves should provide a definitive attendance so long as no answer
sheets are misplaced. Two measures were routinely employed to ensure confidence:
at least two staff members in each lecture theatre would undertake a manual
headcount and, separately, the students were asked to scan their registration
cards as in a normal lecture. In 2007-08 processing the workshop-test data was
possible within 30 minutes of the end of each test for a class of 160 students.
By the start of academic session 2009-10, this task was routinely carried out in
the same timescale but for a class size of 240 students. 

\subsubsection{Workshop-test Marking}
The marking of the workshop tests was done by {\it form recognition} on the scanned
answer sheets. Mostly, this was a problem-free process so long as students used
a dark enough pen or pencil to shade in the answer bubbles and did not change
their mind after shading in an answer. Because the computer software was
programmed to expect a single answer and query both blanks and multiple answers,
it would stop processing the answer sheets and request user intervention for a
decision about the intention of the student in such cases; this happened for
about three or four forms at every test. 

After the data had been downloaded, individual student records were compared
with the nominally correct answers. Thus a final score could be provided to
individual students and, more importantly, the statistical distribution of
answers for individual questions could be checked. If it appeared that a
significant fraction of the class had answered any question wrongly, further
investigations sometimes revealed either a question that was too difficult for
the level of the class or an ambiguously worded question. In all cases, feedback
was given to the lecturer who had set the question so that subsequent questions
would be refined.

Overall, the multiple-redundancy involved in checking the marks ensured complete
confidence in the accuracy of the data and, although these checks could
typically be completed within 30-60 minutes of the test finishing, publication
of the test results was typically delayed for several days to allow further
quality assurance checks to be undertaken. 

\section{Results and Analysis}
The introduction of computer-based technologies to aid the administration of the
first year physics class has proved extremely beneficial. The main objectives of
monitoring attendance and tracking students identified as being at risk of
disengaging from study hinged on the production of an accurate class register
early in the session. Although computer-based technologies assisted in the
production of an initial class register, it was found that significant human
effort was required to finalize it. Subsequently, confidence in the accuracy of
data collated as well as turnaround times for processing were dramatically
improved over previous years when computer-based technologies were not used
(Table~\ref{tab1}).

\begin{table}[ht!]
\begin{center}
  \begin{tabular}{ |l|c|c|c|}
    \hline
    Task 				&2006-07		&2007-08			&2009-10\\ 
					&$\sim140$\,students	&$\sim160$\,students		&$\sim240$\,students \\ \hline \hline
    Produce class register		&3-5\,weeks	&3\,weeks	&3\,weeks \\ \hline
    Take lecture attendance		&N/A		&Weekly		&Daily \\ \hline
    Take laboratory attendance		&Submitted work &60\,minutes	&30\,minutes \\ \hline
    Take workshop-test attendance	&N/A		&180\,minutes	&30\,minutes \\ \hline
    Mark workshop-test scripts		&N/A		&180\,minutes	&30\,minutes	\\ \hline
  \end{tabular}
\caption{Data showing improvements in turnaround times for administration of
tasks in the first year physics class.}
\label{tab1}
\end{center}
\end{table}

The first column shows representative data from 2006-07, before which
computer-based technologies were not in use. Thus, lecture attendance was not
taken before 2007-08 and the workshop-tests were not part of the continuous
assessment in the class. Historically, the laboratory attendance was derived
from the submitted work at the end of each semester by which point there was no
possibility for students to catch up on missed work. In 2007-08, with the
introduction of the computer-based technologies, turnaround times for data
processing and, particularly, some marking tasks decreased significantly. By
2009-10, the system was adequate for coping with the unexpected 50\% increase in
class size. The turnaround times for production of a class register have not
improved because this process is primarily driven by flux in the student
population near the start of the session.

The decrease in turnaround times and associated increase in confidence regarding
the accuracy of the data collected has allowed issues of absenteeism in the
student body to be addressed at an early stage in the semester rather than at
the end. In the case of the workshop-tests and laboratory classes, students can
be contacted within 30-minutes of the end of a workshop-test or the start of a
laboratory. Early contact of absent students has proved vital in terms of
improving student retention and progression statistics. Indeed, it is believed
that this is a contributory factor to the $\sim13\%$ increase in direct progression
from first to second year physics between 2006-07 and 2008-09. 

The {\it form recognition}-based marking of the workshop-tests has proved invaluable
in terms of providing students with feedback on their progress. Staff are now in
a position to mark and return assessment feedback within 30-minutes of the end
of an assessment session for a class size of 240 students. In the previous
technology-sparse era, this would not have been possible or even considered due
to the numbers of students involved. 

\section{Summary and Future Work}
The main aims in streamlining the administration of the first year
physics class appear not only to have been effective but, on some levels,
surpassed. The almost immediate processing of lecture and, more importantly,
laboratory and workshop-test attendance data with previously unknown levels of
confidence has allowed the early identification and contact of students
identified as being at risk of disengaging from the process of studying.

 It is
believed that the combination of administration and academic changes have been
contributing factors in helping improve the course pass-rate with $\sim10\%$ and the
first to second year progression rate by $\sim13\%$ between 2006-07 and 2008-09.
Experience gained in 2007-08 was further developed and refined in 2008-09 so
that the unexpected 50\% increase in student numbers by the start of the 2009-10
session did not present significant problems in class administration. 
Future work will centre around rolling out these techniques to other large
classes within the School specifically the second year physics class,
currently also undergoing a surge in student numbers. Interest in the
administration techniques developed herein is also being shown further afield
within the University of Glasgow.

\section*{Acknowledgements}
The authors would like to thank Dr Stephen McVitie, current class head of the
first year physics course at the University of Glasgow, for many helpful
discussions in refining the logic of these new processes and also for
proof-reading this paper prior to submission.  The authors would also like to
thank Mr Ignacio Santiago Prieto at the University of Glasgow for the providing
the Spanish translation of the title, abstract and keywords.

\newpage
\section*{References}
\bibliography{mmcbiblio}


\end{document}